\documentclass[unsortedaddress,aps,preprint]{revtex4} 
\usepackage{graphicx}
\usepackage{dcolumn}
\usepackage{amsmath}
\usepackage{amssymb}
\usepackage{amsfonts}
\usepackage{bm}
\usepackage{color}
\usepackage[normalem]{ulem}
\usepackage{hyperref}
\hypersetup{colorlinks=true,linkcolor=blue,urlcolor=magenta}
\newcommand{\be}{\begin{equation}}
\newcommand{\ee}{\end{equation}}
\newcommand{\ba}{\begin{eqnarray}}
\newcommand{\ea}{\end{eqnarray}}

\begin{document}
\title{Self-assembled structures of colloidal dimers and disks on a spherical surface}
\author{Nkosinathi Dlamini$^1$, Santi Prestipino$^2$, and Giuseppe Pellicane$^{1,3,4}$\footnote{Email: {\tt gpellicane@unime.it}}}
\affiliation{$^1$School of Chemistry and Physics, University of Kwazulu-Natal and National Institute of Theoretical Physics (NIThEP), 3209 Pietermaritzburg, South Africa \\$^2$Dipartimento di Scienze Matematiche ed Informatiche, Scienze Fisiche e Scienze della Terra, Universit\`a degli Studi di Messina, Viale F. Stagno d'Alcontres 31, 98166 Messina, Italy \\$^3$Dipartimento di Scienze Biomediche, Odontoiatriche e delle Immagini Morfologiche e Funzionali, Universit\`a degli Studi di Messina, 98125 Messina, Italy \\$^4$CNR-IPCF, Viale F. Stagno d'Alcontres, 98158 Messina, Italy}

\begin{abstract}We study the self-assembly on a spherical surface of a model for a binary mixture of amphiphilic dimers in the presence of guest particles via Monte Carlo (MC) computer simulation. All particles have a hard core, but one monomer of the dimer also interacts with the guest particle by means of a short-range attractive potential. We observe the formation of aggregates of various shape as a function of the composition of the mixture and of the size of guest particles. Our MC simulations are a further step towards a microscopic understanding of experiments on colloidal aggregation over curved surfaces, such as oil droplets.
\end{abstract}
\maketitle
\section{Introduction}

Colloidal particles dispersed in a fluid medium are widely considered to be an ideal system where self-assembly can be explored, since they can be resolved and tracked in real time using optical microscopy~\cite{babic:2016}. Aggregation of colloidal particles is often the outcome of steric stabilization by electrostatic repulsion, which is achieved by modifying the salt concentration or by adding chemicals as stabilizing agents --- like in the case of gold colloids~\cite{weitz:1984} or silica and polystirene particles~\cite{lin:1989,lin:1990}. The morphology of colloidal aggregates depends on the prevailing aggregation mechanisms and on the particle shape~\cite{babic:2016}, and is typically observed in ramified or compact clusters of fractal dimensionality~\cite{gonzalez:2004}, in a number of crystalline and amorphous solids~\cite{xu:2016}, and in mesophases~\cite{yang:2018,hagan,prestipino4}, which are partially ordered phases that are intermediate between liquids and crystals (e.g., cluster fluids, liquid crystals, and quasicrystals). Colloidal nanocrystals are even able to self-assemble in crystalline superlattices with an intricate structure~\cite{boles:2016}.

In the last few decades, many researchers have focused on the self-assembly of colloidal particles at an interface, which may serve as a scaffold or template for particles' aggregation. Assembly at air-liquid and liquid-liquid interfaces is driven by a complex interplay of entropic and enthalpic forces~\cite{thapar:2015}. The ability of oil-water interfaces to trap micron-sized particles has been known for over a century~\cite{ramsden:1904,pickering:1907}, and the strong binding of colloidal particles to fluid interfaces (the binding energy being even thousand times stronger than thermal energy) is also evidenced by the stabilization of foams and emulsions against decomposition~\cite{evans:1999,morrison:2002}. Self-assembly of colloidal particles on a flat surface can only rely on the control of interparticle interactions at the interface~\cite{pieranski:1980,hurd:1985,onoda:1985,ruan:2007,retsch:2009}. However, thanks to recent progress in microfluidics it has become possible to modify the interfacial geometry to trap colloidal particles, thus extending the initial range of applications of colloidal self-assembly. Indeed, curved phases of matter are found in a large number of systems, including biological entities like cells and viral capsids, and the competition between the tendency to self-assemble and the geometric frustration originated by the curved interface can lead to novel structures, which are simply impossible to obtain over flat interfaces~\cite{bowick:2009}). Recently, spherical boundary conditions are starting to be employed also in the different realm of ultracold quantum particles, where ``phases'' with polyhedral symmetry are expected~\cite{prestipino4} and Bose-Einstein condensation has peculiarities that are well within reach experimentally~\cite{tononi}.

Concrete realizations of spherical crystals are found in emulsions of two immiscible fluids, such as oil and water, which are stabilized against droplet coalescence by coating the interface of one of the fluids with small colloidal particles~\cite{sacanna:2007}. A non-trivial issue is overcoming the strong binding of colloidal particles to the (liquid-liquid) interface and allowing them to diffuse quickly enough, i.e., like in a true fluid, to facilitate self-assembling over the substrate. Recently, that was achieved by very efficient functionalization with complementary DNA strands of both the surface of oil droplets (which was stabilized with sodium dodecyl sulphate (SDS), i.e., a micelle-forming surfactant) and the surface of colloidal particles~\cite{joshi:2016}. A fluid-like diffusion was reached by allowing colloidal particles to anchor on rafts of polylysine-g[3.5]-polyethylene glycol-biotin (PLL-PEG-bio), which are free to slide on the surfactant. Upon increasing the concentration of SDS, the colloidal particles attached to the surface have been observed to undergo aggregation as a result of the depletion effect driven by the excluded area of the surfactant micelles~\cite{joshi:2016}.

Spherical droplets have also been coated with polystyrene latex particles~\cite{velev:1996} to form aggregates with a rigid shell, called colloidosomes in analogy to liposomes~\cite{dinsmore:2002}. Structures resulting from the encapsulation of colloidal particles~\cite{munao1} are potential candidates for the delivery of drugs and vaccines, and may be used as vehicles for the slow release of cosmetic and food supplements.

In general, self-assembly of colloidal particles on a spherical surface is an important paradigm to understand the structuring of membrane cells, which exhibit stable domains. The distribution and composition of these domains over the cell surface determines the interaction energy between different cells, ultimately driving the organization of the crowded environment inside biological organisms, where a huge number of cells is present~\cite{bott:2016}.

Recently, we have studied in 3D space~\cite{prestipino1} and on a plane~\cite{prestipino2} an implicit-solvent description of a dispersion of two colloidal species --- an amphiphilic dimer and a guest spherical/circular particle --- where the smaller monomer in the dimer is solvophobic and has a strong affinity for the guest particle. In this paper, we consider the same system embedded in a spherical surface. By establishing bonds with two nearby curved disks, the smaller monomer provides the glue that keeps the disks together, which is the mechanism by which disks can form aggregates. However, once an aggregate of disks has been covered with dimers, a further growth of the aggregate is obstructed by the steric hindrance of the coating shell. Since the dimer-disk attraction is of limited range, mostly zero- (``micelles'') and one-dimensional aggregates (``chains'') are expected to form for moderately low amount of disks, while two-dimensional self-assembled structures (i.e., stratified lamellae) can possibly occur under equimolar conditions.

The paper is organized as follows. In the next Section we describe the model and the method employed. In Section 3 we present and discuss our results. Finally, we report our conclusions in Section 4.

\section{Model and method}

The investigated model is the curved-surface analog of the same mixture of dimers and spherical guest particles that has been studied in Refs.~\cite{prestipino1,prestipino2,munao2}: a dimer consists of two tangent hard calottes (i.e., disks following the surface of the sphere) with curved diameters $\sigma_1$ and $\sigma_2=3\sigma_1$, whereas guest particles are represented as hard calottes of size $\sigma_3$ (below, we generically refer to these particles as ``disks''). In addition to impenetrability of all particle cores, we put an attraction between the small monomer and the disk, modeled as a square-well potential of depth $\epsilon$; the width of the well is set equal to $\sigma_1$. In the following, $\sigma_2$ (i.e., the diameter of the large monomer) and $\epsilon$ are taken as units of length and energy. Finally, we call $N_1=N_2$ and $N_3$ the number of dimers and disks, respectively; hence, $N=N_1+N_3$ is the total number of particles and $\chi=N_3/N$ is the (disk) composition. 

Most of the data have been collected for a fixed number $N_3=400$ of disks with diameter $\sigma_3=\sigma_2$ and varying composition. The number density is $\rho^*\equiv(N/A)\sigma_2^2=0.05$ ($A$ being the area of the spherical surface), but we have also performed a few simulations for $\rho^*=0.25$ to probe the regime of moderately high curvatures. We have analyzed the system behavior for a number of compositions: $\chi=20\%,33\%,50\%$, and $80\%$. Once $N$ and $\chi$ have been set, the number of dimers follows accordingly (notice that the sphere radius $R$ is uniquely determined from $N$ and $\rho^*$). We also examine how self-assembly changes when the disk diameter is increased up to $5\sigma_2$.

Simulations are carried out using the standard Metropolis algorithm in the canonical ensemble. Typically, a few hundred million Monte Carlo (MC) cycles are performed, one cycle consisting of $N$ trial moves. Both translational and rotational moves are carried out for dimers. In performing a translational move, the midpoint of the arc joining the monomer centers is randomly shifted on the sphere, while keeping the direction of the subtended chord fixed in the embedding three-dimensional space. In a rotational move, it is the midpoint of the arc between the monomers that is fixed, while the chord is rotated at random. The maximum random shift and rotation are adjusted during the equilibration run so as to keep the ratio of accepted to total number of moves close to 50\%. The schedule of each move is designed so that detailed balance holds exactly. Particles are initially distributed at random on the sphere (using a variant of the Box-Muller algorithm~\cite{Krauth}). Then, the system is quenched to $T^*\equiv k_BT=0.15$ or $0.10$ and subsequently relaxed until some stationary condition is established. We have checked that cooling the system slowly in steps, starting at each temperature from the last configuration produced at a slightly higher temperature, does not make any substantial difference in the structural properties of the steady state. This is due to the fact that the simulated systems are overall dilute.

\begin{figure}
\centering
\includegraphics[width=12cm]{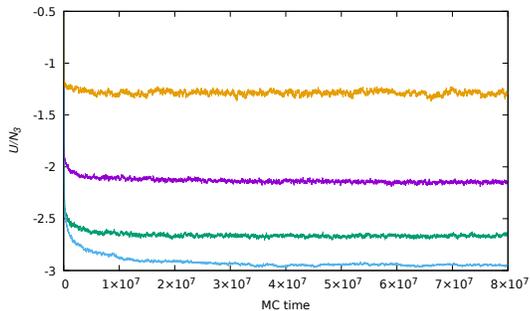}
\caption{Energy evolution, as a function of Monte Carlo time, for $\sigma_3=\sigma_2,T^*=0.10,\rho^*=0.05$, and various disk compositions (from top to bottom, $\chi=20\%,33\%,50\%$, and $80\%$).}
\end{figure}

A property signalling how far the system is away from equilibrium is the total potential energy $U$: an energy fluctuating around a fixed value for long is the hallmark of (meta)stable equilibrium. In $\epsilon$ units, $U$ gives the total number of 1-3 contacts in the current system configuration. Hence, a stationary value of $U$ indicates that aggregates have eventually reached a nearly stable structure. Typically, $10^8$ cycles suffice for reaching a stationary state of low density. This is clearly illustrated in Fig.\,1, showing the energy evolution as a function of Monte Carlo cycles for $T^*=0.10,\rho^*=0.05$, and a number of compositions. Once equilibrium (or a steady state whatsoever) has been reasonably attained, we gain insight into the nature of aggregates mainly by visual inspection. We also compute the radial distribution function (RDF) of disks, $g_{33}(r)$, in a rather long production run of $10^7$ cycles (we have checked that statistical errors on the RDFs are indeed negligible). For the sake of comparison, similar studies of 2D binary mixtures at considerably higher total density are executed for one order of magnitude less MC steps~\cite{fiumara}.  Even in a strongly heterogeneous system where mesoscopic structures are present, $g_{33}(r)$ bears valuable information on the arrangement of disks in the close neighborhood of a reference disk. Two disks are said to form a bound pair when their distance is not larger than $r_{\rm min}=\sigma_3+3\sigma_1$~\cite{prestipino2}: this is the maximum distance at which two disks can still be in contact with the same small monomer (exactly placed in the middle). For $\sigma_3=\sigma_2$, this implies that a disk makes bonds with all its first and second neighbors, identified as such through the RDF profile.

\section{Results}

In the following, we first comment on the simulation results for a mixture of disks and large monomers having the same size (Section 3.1). We consider systems of both low density ($\rho^*=0.05$) and moderate density ($\rho^*=0.25$). Next, we will see what changes when dimers are much smaller than disks (Section 3.2). By visual inspection we can easily ascertain the nature of the structures present in the stationary configurations of the low-temperature system.

\subsection{Same size of dimers and disks}

\begin{figure}
\centering
\includegraphics[width=7cm]{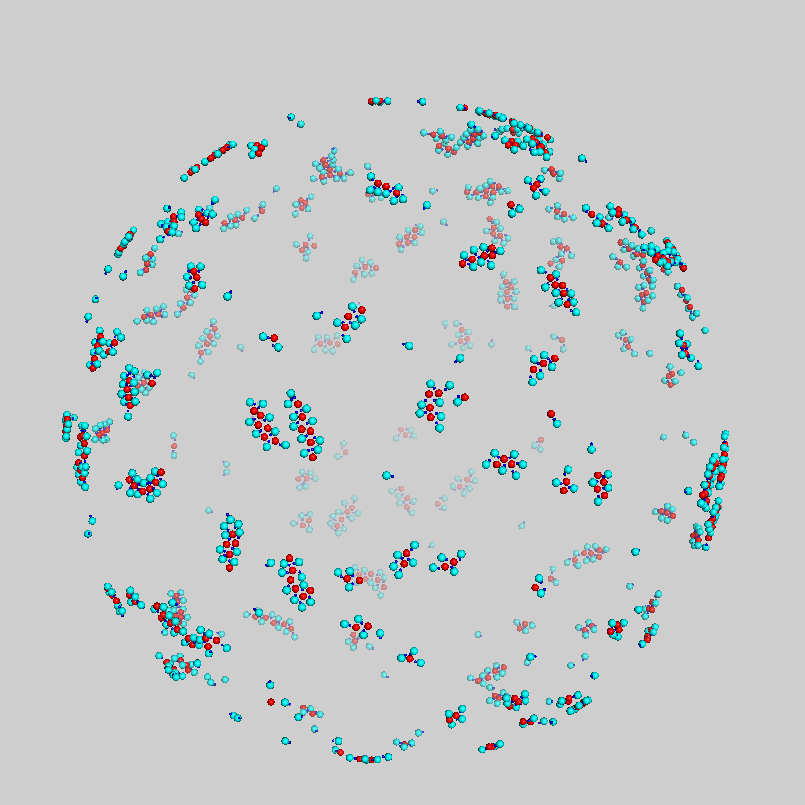}
\includegraphics[width=7cm]{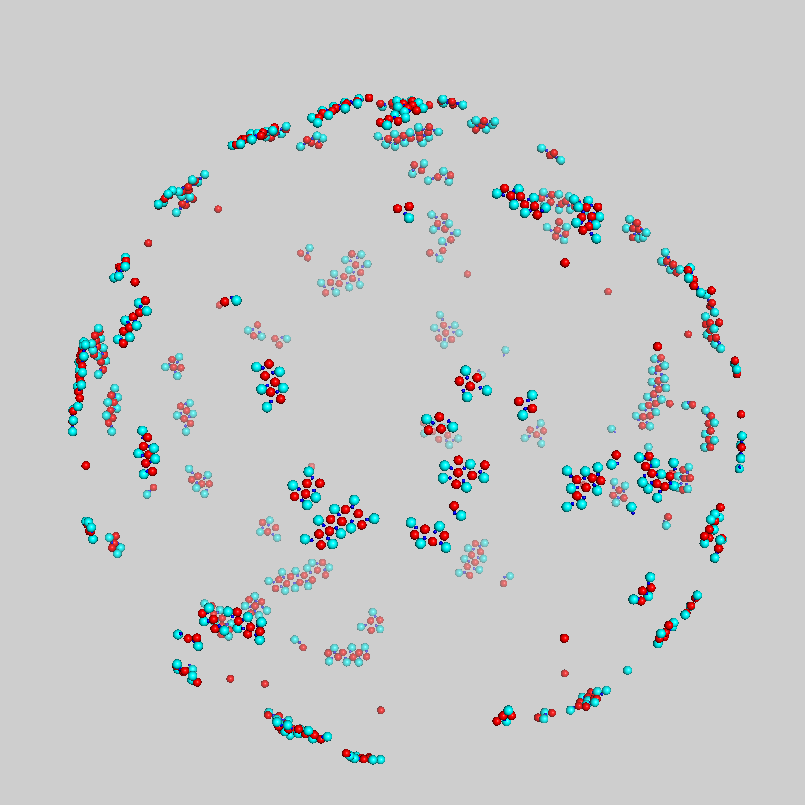}
\caption{Typical configuration of the mixture for $\sigma_3=\sigma_2$ and $\rho^*=0.05$, after a long run at $T^*=0.15$. The snapshots refer to a system of composition $\chi=33\%$ (left) and $\chi=50\%$ (right).}
\end{figure}

We initially set the density equal to $\rho^*=0.05$ and the disk size to $\sigma_3=\sigma_2$. For $T^*=0.15$, the equilibrated system is a fluid of small globular clusters for all disk compositions, see examples in Fig.\,2. Only for values of $\chi$ lower than about 10\%, dimers form a well-definite coating shell around disks. Thermal fluctuations for $T^*=0.15$ are still too important to allow for the formation of more elaborate structures.

\begin{figure}
\centering
\includegraphics[width=7cm]{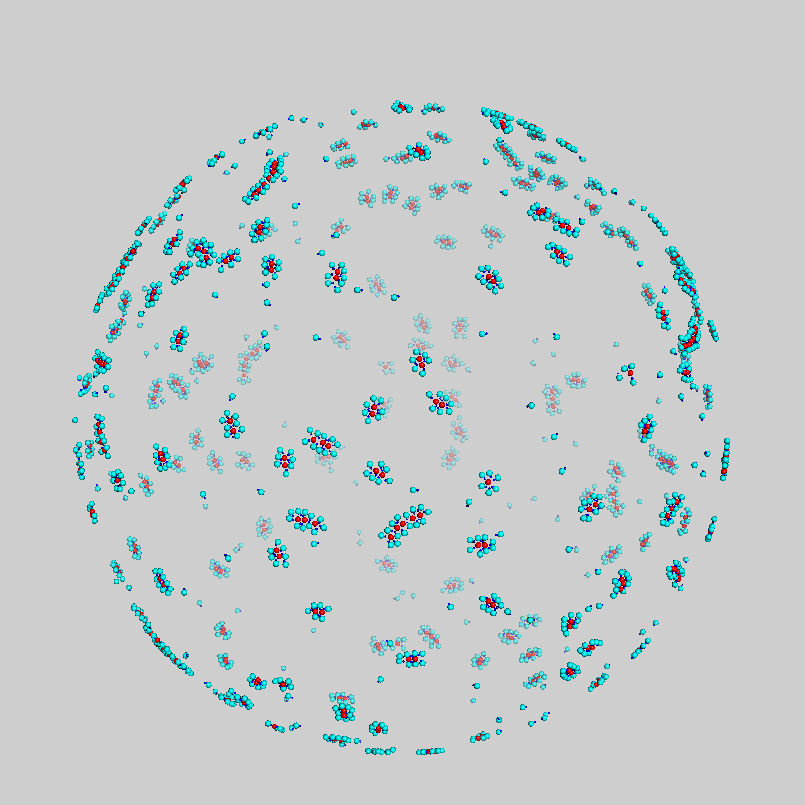}
\includegraphics[width=7cm]{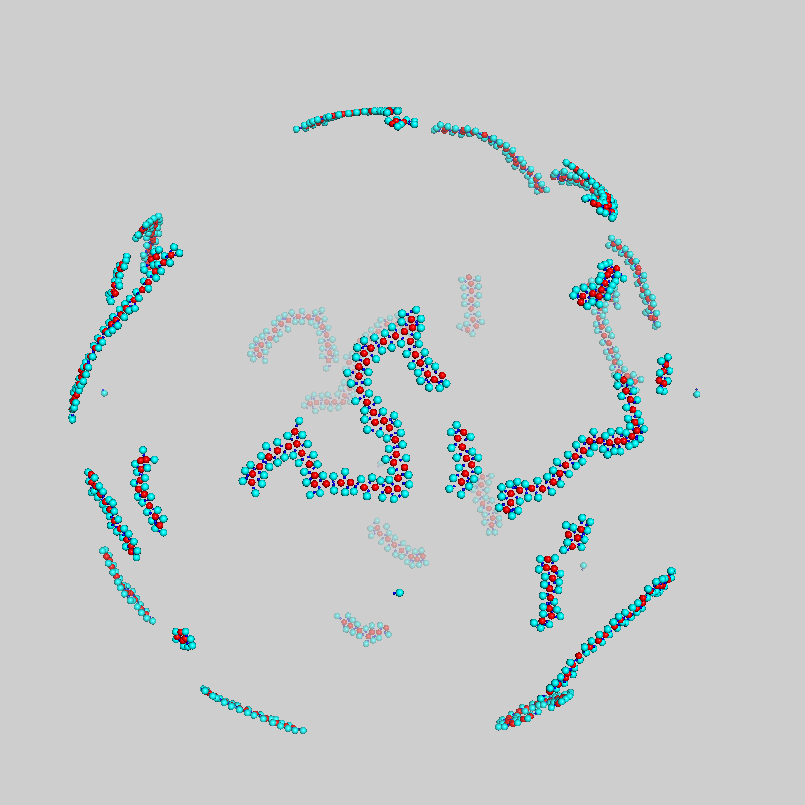}
\includegraphics[width=7cm]{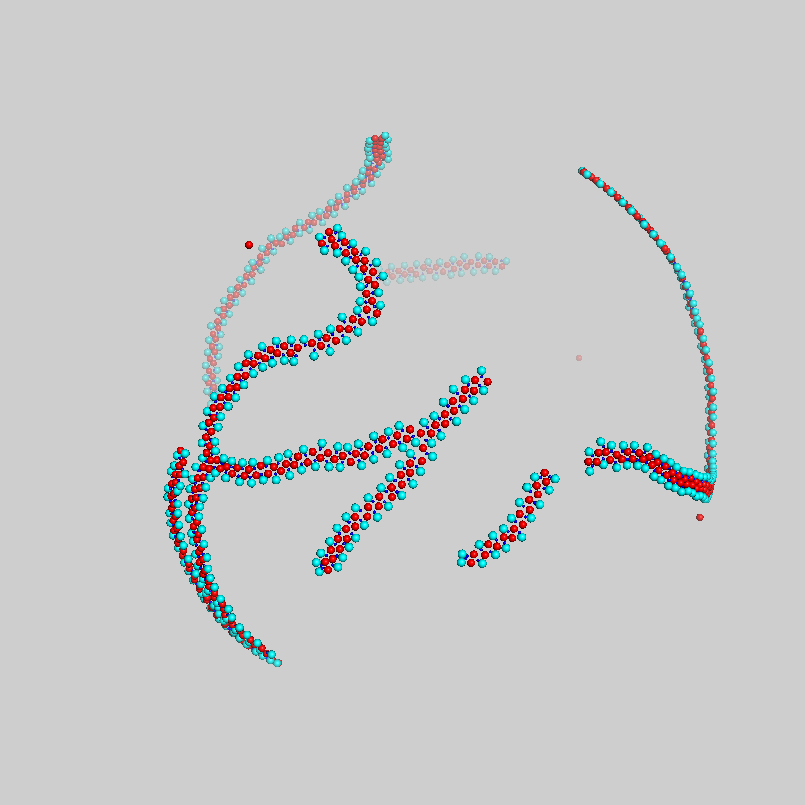}
\includegraphics[width=7cm]{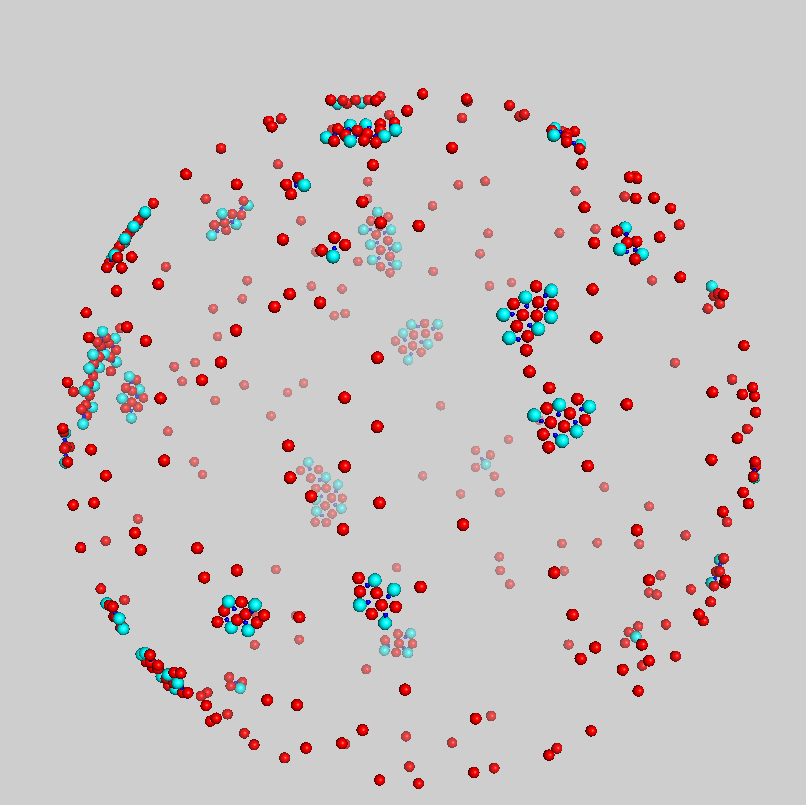}
\caption{Typical configuration of the mixture for $\sigma_3=\sigma_2$ and $\rho^*=0.05$, after a long run at $T^*=0.10$. The snapshots are taken at different compositions: $\chi=20\%,33\%,50\%$, and $\chi=80\%$ (from top left to bottom right).}
\end{figure}

Things change radically for $T^*=0.10$, where the nature of aggregates is more varied. For $\chi=20\%$ or lower we invariably observe small groups of disks surrounded by dimers (Fig.\,3, top left panel), while for higher compositions up to $50\%$ aggregates are more elongated and worm-like (see top right and bottom left in Fig.\,3). A closer look at such ``worms'', which are obviously the 2D analog of lamellae, reveals that they are assembled from a repeating unit, like a polymer chain. For still higher $\chi$, the mean size of aggregates returns to being small again since the number of gluing dimers is insufficient for all disks and many disks then remain unbound (Fig.\,3, bottom right panel). Therefore, aggregates only achieve large sizes when the number of disks roughly matches that of dimers.

The dynamics of aggregation in the present model is easy to explain. Initially, when the system is still disordered, aggregation of disks proceeds very fast through the formation of bonds between disks and dimers. As an aggregate grows in size, however, its surface becomes increasingly rich in large monomers, which are inert particles; eventually, an aggregate stops growing when its disks and small monomers all lie buried under the surface. While local adjustments of the structure still occur at a high rate, the merging of two disconnected aggregates (or the breaking of a long chain) is highly suppressed and only takes place on much longer time scales. The existence of two regimes of aggregate growth (fast and slow), corresponding to a transition from diffusion-limited to reaction-limited aggregation~\cite{lin:1989}, is reflected in the crossover of $U$ from an exponential to a sub-exponential decay (as evidenced in Fig.\,1).

For $T^*=0.10$, the mechanism underlying the structure of aggregates is mainly energy minimization, whereas entropy considerations would play a minor role. However, entropy is decisive in shaping the large-scale distribution of aggregates on the sphere, inasmuch as their structure maintains a certain flexibility (see more below). Clearly, also the relative size of particles and the range of 1-3 attraction are important. As already commented, it is the short-range character of the attraction that is responsible for the essentially one-dimensional geometry of the larger aggregates.

\begin{figure}
\centering
\includegraphics[width=7cm]{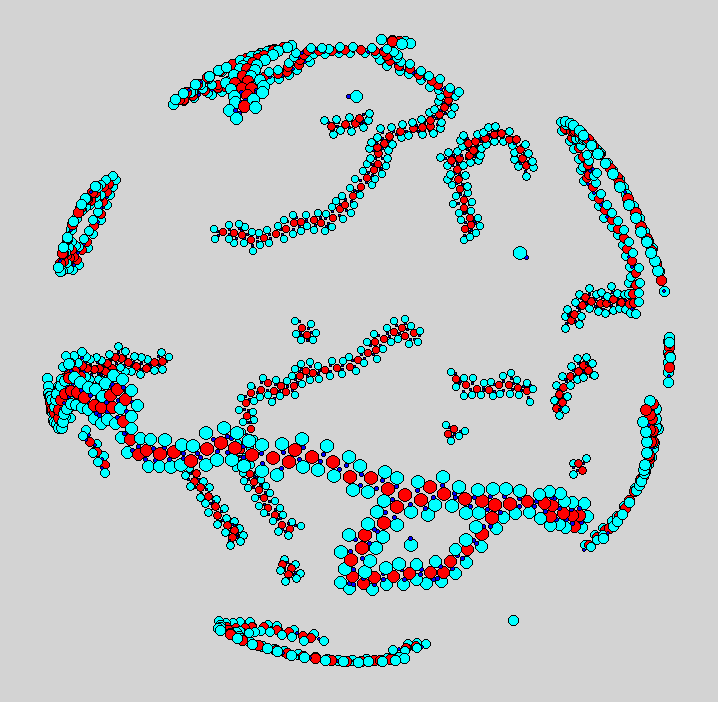}
\includegraphics[width=7cm]{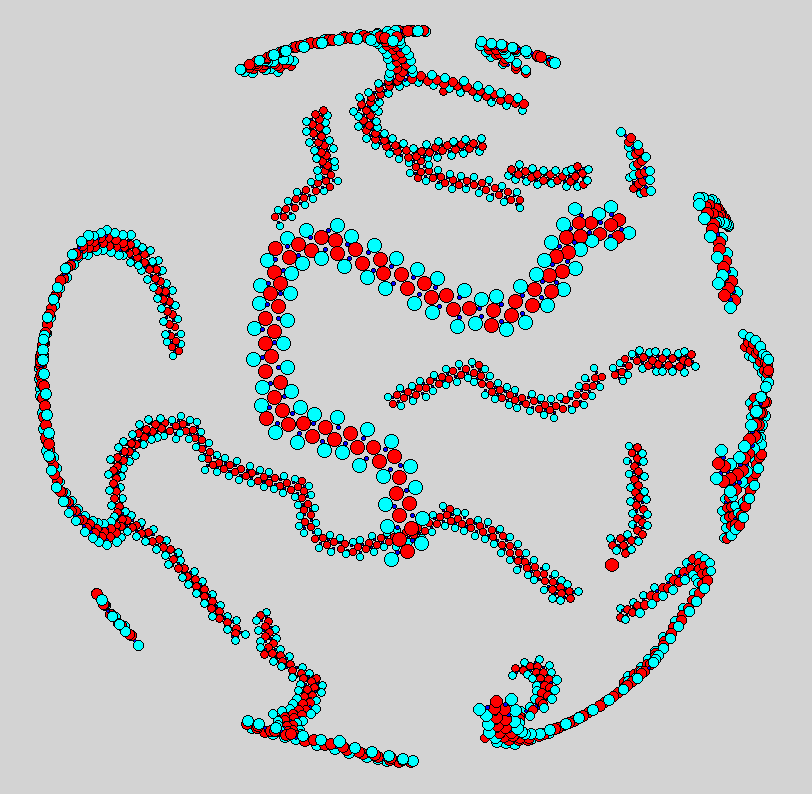}
\caption{Typical configurations of the mixture for $\sigma_3=\sigma_2$ after ``only'' $3\times 10^7$ MC cycles. The snapshots are taken for $T^*=0.10$ and $\rho^*=0.05$, and refer to $\chi=33\%$ (left) and $\chi=50\%$ (right).}
\end{figure}

Particularly interesting are the systems with $\chi=33\%$ and $\chi=50\%$, which we see enlarged in Fig.\,4. Here most of the aggregates are flexible worms, i.e., chain-like aggregates with a small bending modulus. The geometry of the worm backbone is dictated by the necessity to keep the energy as small as possible at the given composition: this is accomplished by a straight chain of disks for $\chi=33\%$ (Fig.\,4, left panel) and by a zig-zag chain for $\chi=50\%$ (Fig.\,4, right panel). Both chain morphologies allow disks to bind all dimers, so that in the long run no free particles would be left in the box. Occasionally, a worm bends to the point that a closed loop appears --- see the example on the left in Fig.\,4.

\begin{figure}
\centering
\includegraphics[width=7cm]{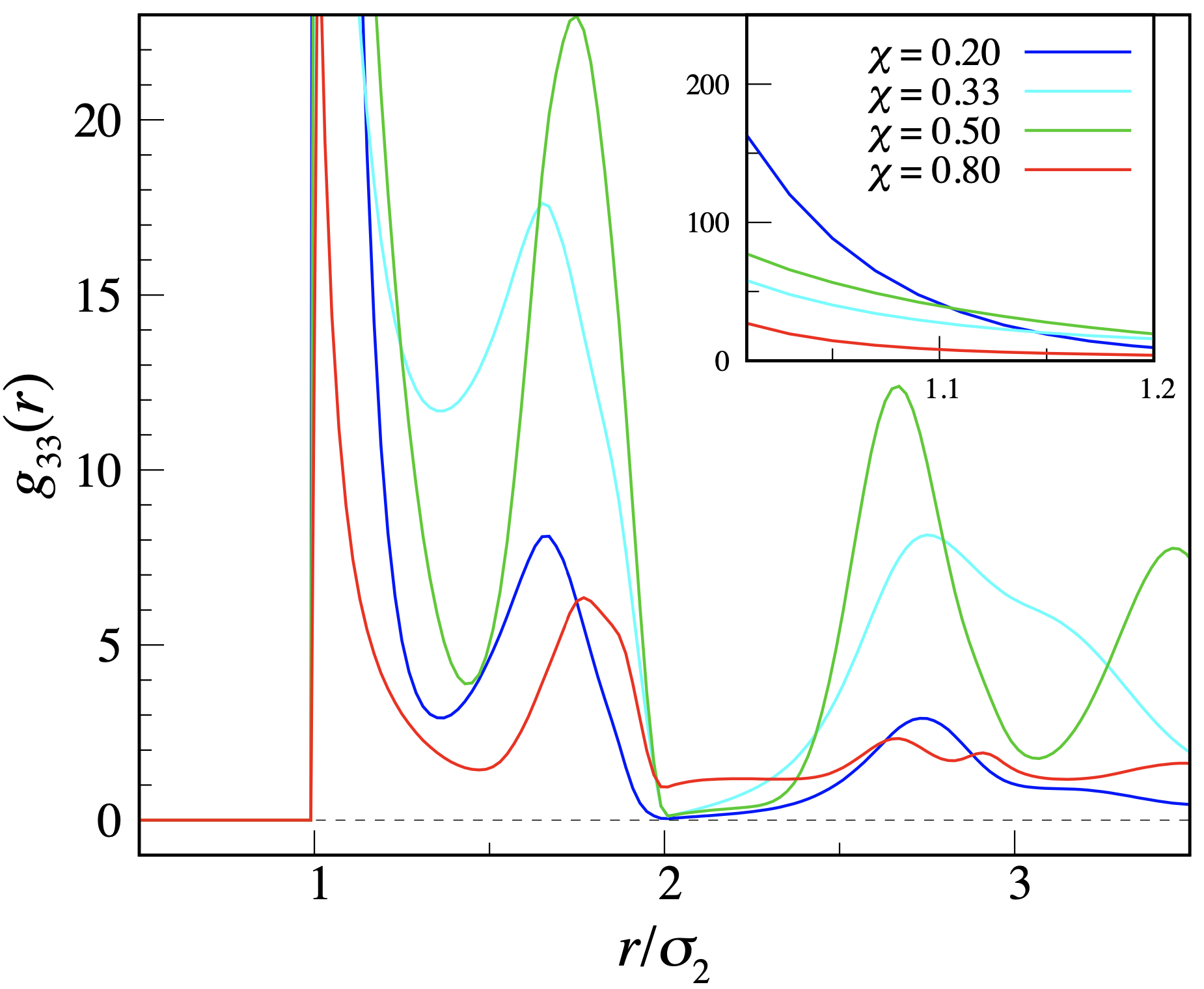}\,\,\,\,\,\,
\includegraphics[width=7cm]{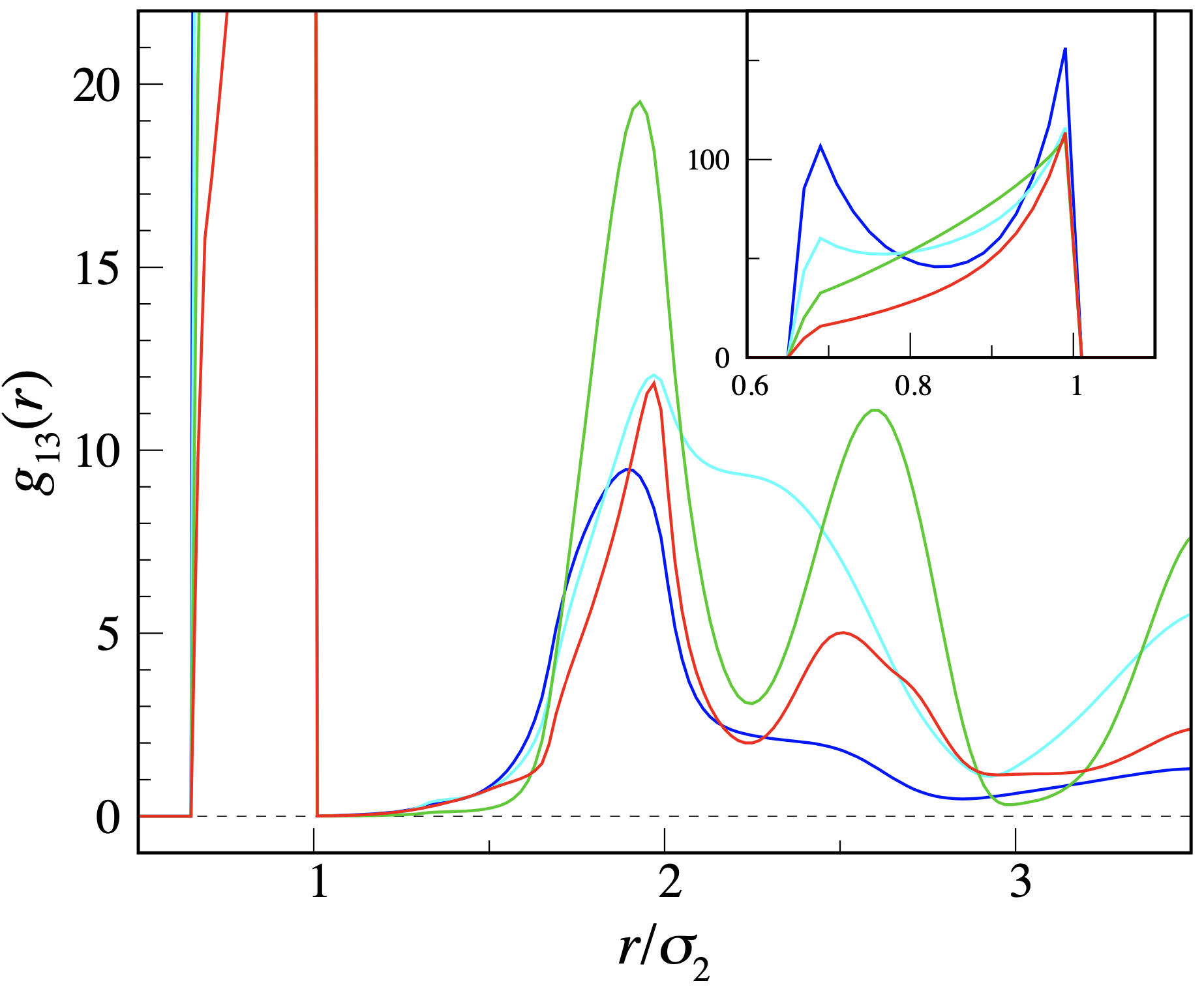}
\caption{Mixture of dimers and disks with $\sigma_3=\sigma_2$ after $10^8$ MC cycles, for $T^*=0.10$ and $\rho^*=0.05$. Left: $g_{33}(r)$. Right: $g_{13}(r)$. In the inset, the short-range structure of the RDFs is highlighted. The color code is the same for both panels, see left-panel inset.}
\end{figure}

In Fig.\,5 we collect the RDFs for various compositions at low temperature ($T^*=0.10$). A large $g_{33}$ value at contact is the most distinct signature of the existence of aggregates of disks. The short-distance structure in $g_{33}$ is richer for intermediate $\chi$ values, where the physiognomy of aggregates is better defined. The rather pronounced second-neighbor peak at $\chi=50\%$ is the result of the zig-zag structure of the chain backbone at this composition. As to $g_{13}(r)$, its short-distance profile is sharper for the lowest compositions, where the number of dimers that are in close contact with the same sphere is higher (see inset of right panel). A high third-neighbor peak for intermediate compositions is the signature of the existence of extended aggregates of dimers and guest disks consisting of a periodically repeating unit.

\begin{figure}
\centering
\includegraphics[width=7cm]{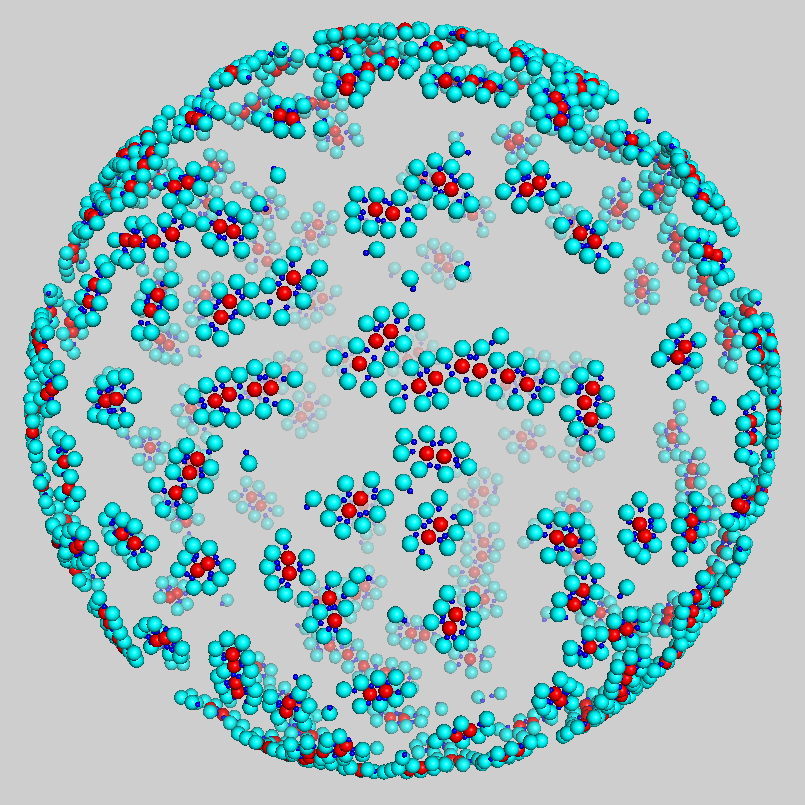}
\includegraphics[width=7cm]{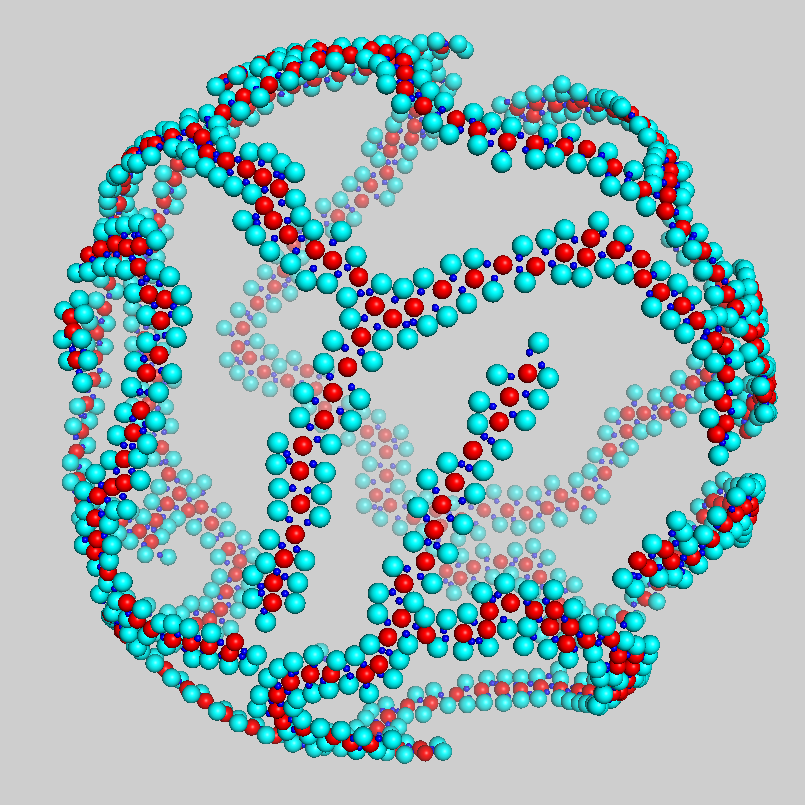}
\includegraphics[width=7cm]{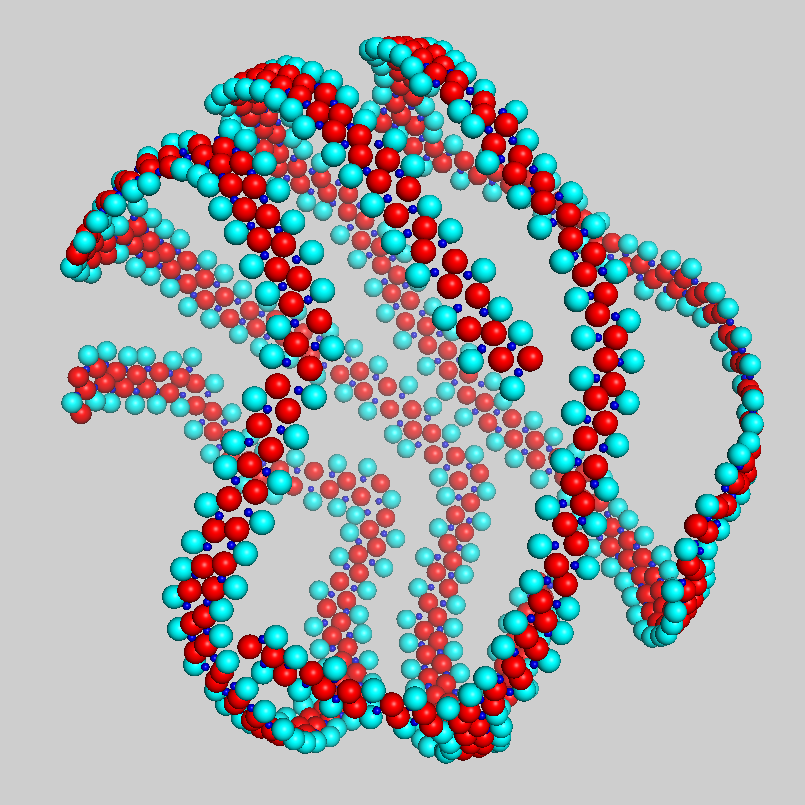}
\includegraphics[width=7cm]{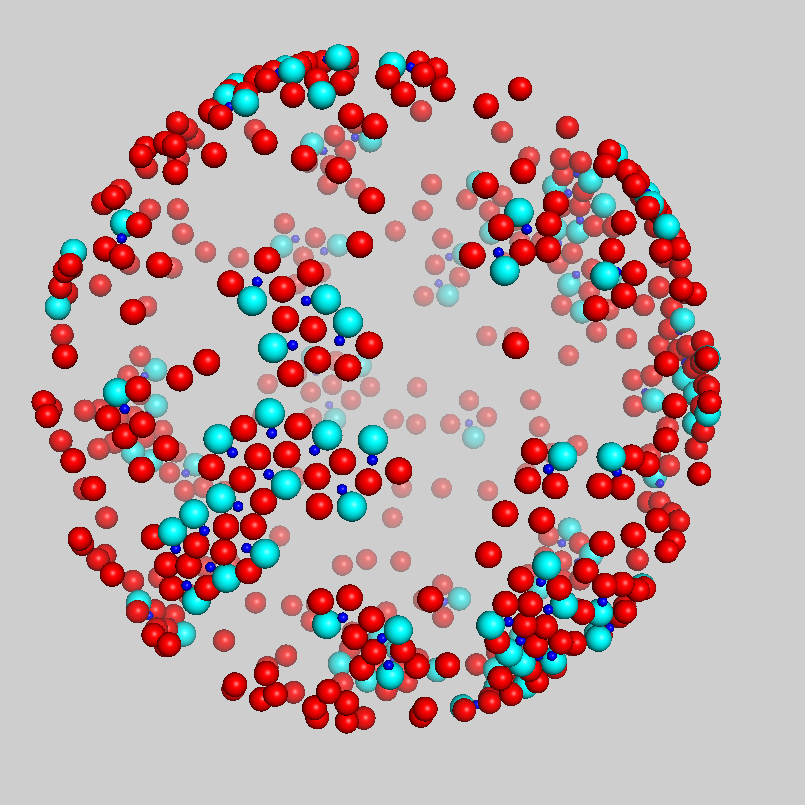}
\caption{Typical configuration of the mixture for $\sigma_3=\sigma_2$ and $\rho^*=0.25$, after a long run carried out at $T^*=0.10$. The snapshots are taken at different compositions: $\chi=20\%,33\%,50\%$, and $\chi=80\%$ (from top left to bottom right).}
\end{figure}

As the density increases, the nature of self-assembly becomes slightly different. We have studied mixtures for $\rho^*=0.25$, while still keeping $\sigma_3=\sigma_2$ and the temperature fixed at 0.10 (see Fig.\,6). When the composition is low the aggregates are slightly elongated capsules, see top left panel of Fig.\,3, like at small density. For $\chi=33\%$, aggregates are definitely chain-like (Fig.\,6, top right panel), but, due to a more crowded environment, they are now joined together in an intricate manner, giving rise to a spanning cluster (i.e., a connected gel-like network encompassing all particles in the system). The onset of an extended network is a remarkable outcome, considering that this structure emerges from very basic interaction rules in a binary system of disks and dimers. At the composition of 50\%, large monomers are more effective in screening a chain from other aggregates and chains then grow much longer. As a result, a chain may wraps a few times around the sphere before merging into another chain (see bottom left panel of Fig.\,6). This is similar to what observed in a one-component system of particles interacting through a SALR (short-range attractive, long-range repulsive) potential~\cite{pekalski}. The latter system is stripe-forming at low temperature; hence, when the particles are constrained to a spherical surface, a stripe may wrap around the sphere, as indeed observed. Finally, at still higher compositions of disks the size of aggregates is again reduced, since there are few disks to bind all dimers and their environment is too crowded to allow aggregates grow long in the early stages of equilibration.

While a further, moderate increase of the density at fixed $N\approx 1000$ would not change much in the above results, we expect that in a huge-density mixture the formation of self-organized aggregates will encounter much difficulty, due to an increased relaxation time and overcrowding of the surface.

\subsection{Increasing the disk size}

\begin{figure}
\centering
\includegraphics[width=7cm]{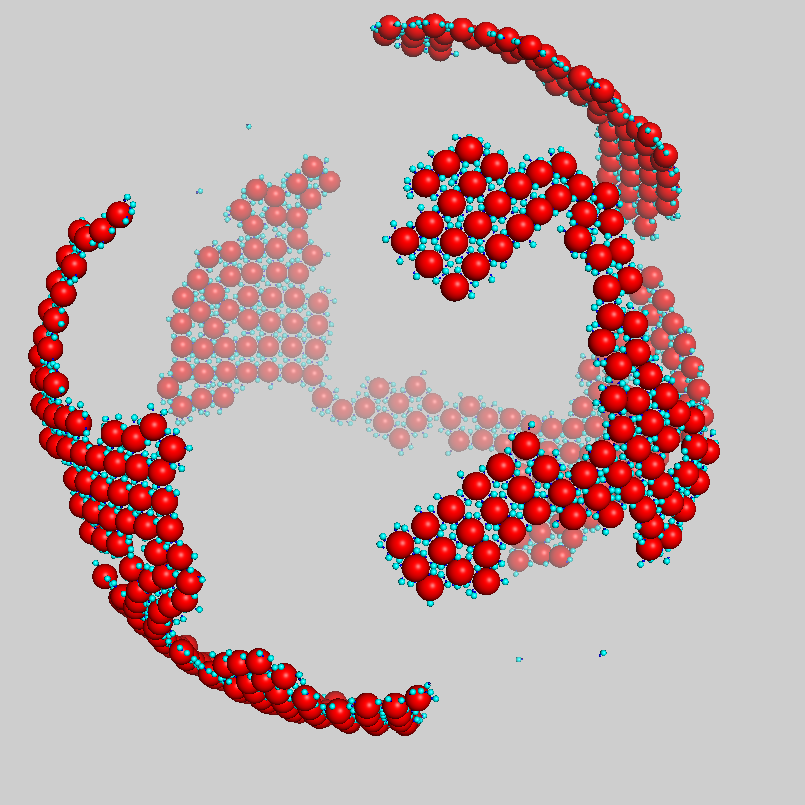}
\includegraphics[width=7cm]{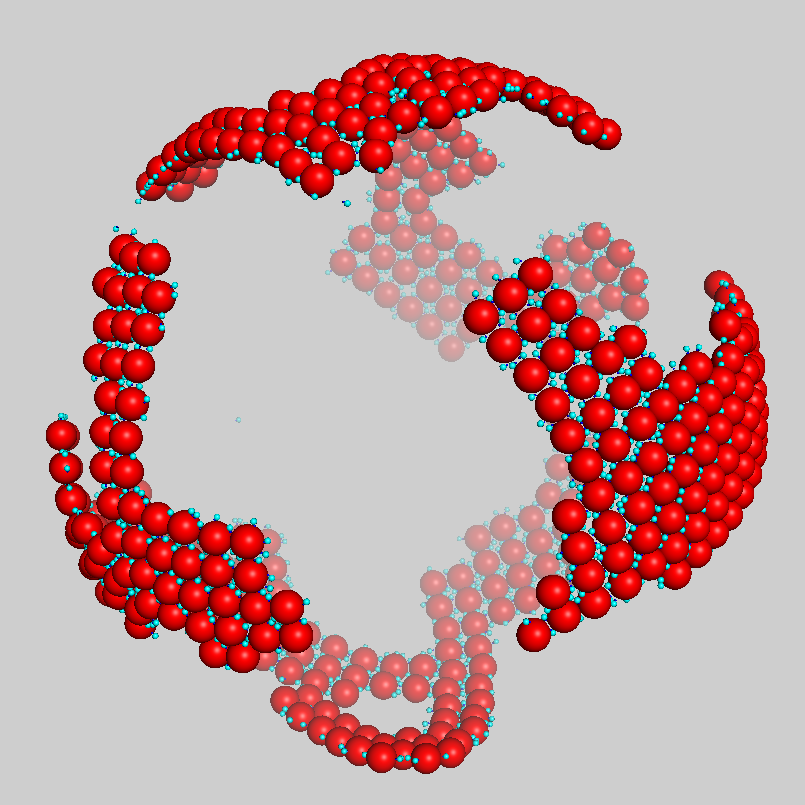}
\caption{Typical configuration of the mixture for $\sigma_3/\sigma_2=4$ (left) and $\sigma_3/\sigma_2=5$ (right), after a long run at $T^*=0.10$. The density of the system is $\rho^*=0.05$ and the disk composition is $\chi=20\%$. The disks are gathered together in patches similar to the tectonic plates floating on the Earth's mantle. The prevailing structural motif is a square of disks, with gluing dimers in the middle interstice.}
\end{figure}

When the size of disks becomes sufficiently large, say four at least, the large monomers fail to adequately screen the attraction between two disk aggregates and, at low temperature, we observe the formation of a condensate over the sphere, in agreement with what found for the system absorbed on a flat interface~\cite{prestipino2}. This is clearly seen in the snapshots reported in Fig.\,7, which correspond to a typical late-time configuration of the system for $T^*=0.10,\rho^*=0.05$, and $\chi=20\%$. For $\sigma_3/\sigma_2=5$, disks mostly occur in the form of a square-symmetric  polycrystalline structure, held together by dimers interspersed between the disks (we count an average of four dimers in each square center, in accordance with the overall disk composition). The prevalent square order of the system is transparent in the profile of $g_{33}(r)$, where the second and third peaks fall at distances that are in a ratio of $\sqrt{2}$ and 2, respectively, with the location of the first peak. It is evident from Fig.\,7 that the size of patches with a clear square motif is nonetheless limited, and the reason for this is twofold. For one thing, this may be the result of an incomplete equilibration; relaxation to equilibrium (coarsening) is slower for a crystallizing system where many particles are hosted in a single cell; as a result, the spontaneous elimination of crystalline defects will take much longer than in a one-component crystal of spherical particles. Even though the onset of crystalline order is favored on a sphere by a lowering of the nucleation barrier~\cite{gomez:2015}, this effect is seemingly small at the probed densities where the curvature of the sphere is small as well. On the other hand, perfect square order is inherently frustrated on the sphere, and this places an upper threshold on the size of the ordered patches (this would still be consistent with the existence of a superstructure of patches, akin to the icosahedral superstructure found in dense systems of hard disks on a sphere~\cite{prestipino3}, but we have no evidence for that).

It is gratifying that the regular structure observed in Fig.\,7 finds a correspondence in the crystalline order of DNA-hybridized polystyrene colloids on the surface of oil droplets, as can be seen in the fluorescence micrographs of Ref.~\cite{joshi:2016}. At variance with these triangular-ordered crystalline patches, however, our system is unique in providing a square-symmetric scaffold for absorption of foreign particles on a spherical substrate.

\section{Conclusions}

We have performed Monte Carlo (MC) computer simulations on a sphere of a binary mixture of asymmetric dimers of tangent hard disks and guest particles at low temperature. Guest particles are curved hard disks interacting with the smaller monomer of the dimer through an attractive square-well potential, whose range is the same as the monomer diameter. We have analyzed the effect of changing the density of the mixture (albeit still keeping the system very sparse), the composition, and the size of guest particles, ranging from the size of the larger monomer in a dimer to five times bigger than that. Despite the simplicity of the model, its self-assembly behavior turns out to be quite rich. Indeed, we observe the formation of various metastable aggregates with a prevailing one-dimensional geometry (i.e., chain-like aggregates with a small bending modulus, including a gel-like network), only driven by the short-range attraction, while the large-scale distribution of aggregates can be understood in terms of entropic considerations. For a large size of the guest particle, the small monomer-guest particle attraction can only hardly be screened by the large monomer, thus favoring the formation of thick condensates. These condensates appear as a square-symmetric polycrystal, which is a non-trivial feature of the model considering that the perfect square lattice is inherently frustrated on the sphere. Our results indicate the possibility of building a square-symmetric scaffold for absorption of external particles on a curved surface.

\acknowledgments{This work has benefited from computer facilities at the South African Center for High Performance Computing (CHPC) under the allocation MATS0887 and from computer facilities at UKZN (cluster HIPPO).}

\end{document}